\documentclass{aa}  

\usepackage{natbib}
\usepackage{graphicx}
\usepackage{txfonts}
\usepackage{lipsum}      
\usepackage{lscape}             
\usepackage{placeins}           
\usepackage{soul}                           

\begin{document}

   \title{Carbon in J0815+4729: Casting a shadow of doubt on the interpretation of CH absorption in the most primitive stars}
   \titlerunning{Carbon in J0815+4729}

   \author{Junbo Zhang\inst{1,2}\corrauth{jbzhang@bao.ac.cn}        
        \and Carlos Allende Prieto\inst{2,3}\corrauth{carlos.allende.prieto@iac.es}
        \and Yeisson Osorio\inst{2,4}\email{yo@ing.iac.es}
        \and Jonay I. Gonz\'alez Hern\'andez\inst{2,3}\email{jonay@iac.es}
        \and David S. Aguado\inst{2,3}\email{aguado@iac.es}
        \and Ivan Hubeny\inst{5}\email{ihubeny.astr@gmail.com}
        \and Shuai Liu\inst{1}\email{liushuai@nao.cas.cn}
        \and Jianrong Shi\inst{1,6,7}\email{sjr@nao.cas.cn}
        \and Rafael Rebolo\inst{2,3,8}\email{rrl@iac.es}
        \and Gang Zhao\inst{1,6}\email{gzhao@nao.cas.cn}
        }

   \institute{National Astronomical Observatories, Chinese Academy of Sciences, Beijing 100101, P.R.China
   \and Instituto de Astrofísica de Canarias, 38205 La Laguna, Tenerife, Spain
   \and Universidad de La Laguna, Departamento de Astrofísica, 38206 La Laguna, Tenerife, Spain
   \and Isaac Newton Group of Telescopes, Apto 321, E-38700 Santa Cruz de La Palma, Canary Islands, Spain
   \and University of Arizona, Tucson, AZ, USA
   \and School of Astronomy and Space Science, University of Chinese Academy of Sciences, Beijing 100049, P.R.China
   \and School of Physics and Technology, Nantong University, Nantong 226019, P.R.China
   \and Consejo Superior de Investigaciones Científicas, E-28006 Madrid, Spain
   }

   \date{Accepted August 20, 2026}
 
  \abstract
   {}
   {We revisit the carbon abundance in the primitive iron-poor star J0815+4729 ([Fe/H] $=$ $-$5.49). J0815+4729 is so enhanced in carbon and rich in oxygen that, in addition to the CH and C$_{2}$ molecular lines, weaker neutral atomic transitions are measurable in its spectrum. This provides a unique opportunity to explore the line formation of \ion{C}{i} and \ion{O}{i} transitions in such an extremely iron-deficient star.} 
   {Based on high-resolution spectroscopy and newly constructed atomic models for carbon and oxygen, we present the first investigation into deriving the carbon and oxygen abundances of this hyper-iron-poor star from neutral atomic transitions under non-local thermodynamic equilibrium (NLTE) conditions.}
   {The carbon abundance we derived from \ion{C}{i} lines is A(C)$_{\rm NLTE} = 8.08 \pm 0.07$, which is consistent with that inferred from C$_{2}$ lines (A(C)$_{\rm LTE} = 7.96 \pm 0.08$), but higher by about 0.6 dex than the value from CH lines (A(C)$_{\rm LTE} = 7.47 \pm 0.17$). Since the CH lines are the most commonly used carbon abundance indicator for extremely metal-poor stars, this result casts doubt on the quantitative interpretation of CH absorption in the most primitive stars. We confirm that J0815+4729 is one of the most carbon-enhanced stars known to date, with [Fe/H] $\le$ $-$4.5. We also determined A(O)$_{\rm NLTE}$ as 7.25 $\pm$ 0.11, with NLTE corrections of about $-$0.11\,dex. The carbon-to-oxygen ratio of the star is [C/O] $=$ 1.06 $\pm$ 0.11.}
   {}

\maketitle
\nolinenumbers

\section{Introduction}
Extremely metal-poor (EMP) stars offer critical insights into the early Universe, the first generations of stars, and the processes that shaped subsequent stellar populations. Among the elements studied in these ancient stars, carbon plays a prominent role. For example, [C/Fe] is commonly used as an indicator to separate the different populations, that is, carbon-enhanced metal-poor (CEMP; [C/Fe] $>$ 1.0, sometimes also $>$ 0.7; \citet{Beers2005, Aoki2007, Placco2014, Bonifacio2025}) from more carbon-deficient stars. These reflect two different star formation scenarios. 

The stellar carbon abundance is determined based on various spectroscopic indicators, such as the \ion{C}{i} lines and the lines of the molecular species CH, C$_{2}$, and CO. Only 16 iron-poor stars with [Fe/H] $<$ $-$4.5\,dex have been identified so far, and only seven stars with [Fe/H] $<$ $-$5.0\,dex \citep{Christlieb2002,Frebel2005,Norris2007,Caffau2011,Hansen2014,Keller2014,Allende2015,Bonifacio2015,Caffau2016,Aguado2018a,Aguado2018b,Starkenburg2018,Aguado2019,Nordlander2019,Limberg2025, Chiti2026}. The quantitative evaluation of the C abundance in these primitive stars is typically based on the CH $G$ band using 1D local thermodynamic equilibrium (LTE) model atmospheres, because the CH features are usually strong enough to be detectable in these most iron-poor stars, even for the two carbon-normal stars, UCAC3 215-112497 and GDR3$\_$526285 \citep{Caffau2011,Limberg2025}.
Only for one of these stars were the C$_2$ and CH molecular lines analyzed \citep{Christlieb2002, Christlieb2004}. No study has employed the \ion{C}{i} lines so far because the transitions are weak and reliable H collisional rates for accurately evaluating potential departures from LTE are lacking. 

The source SDSS~J081554.26+472947.5 (hereafter J0815+4729) was first identified in the baryon oscillation spectroscopic survey (BOSS) of the sloan digital sky survey (SDSS) and was confirmed as a carbon-rich hyper-metal-poor star by \citet{Aguado2018b}. Subsequently, \citet{Gonzalez2020} conducted high-resolution spectroscopic follow-up observations, reporting an iron abundance of [Fe/H]~$= -5.49 \pm 0.14$, a carbon enhancement of [C/Fe]~$= 4.49 \pm 0.11$ (from CH lines), and an oxygen enhancement of [O/Fe]~$= 4.03 \pm 0.12$ (from \ion{O}{i} lines). It belongs to the CEMP-no class, and therefore, its abundances likely reflect those of the pristine material polluted with the ejecta from core-collapse supernovae of the first-generation stars \citep{Gonzalez2020, Gonzalez2023, Bonifacio2018}.

It is commonly assumed that the only carbon indicators available in stars of [Fe/H] $<$ $-$4.5 are molecular bands; however, J0815+4729 provides a unique opportunity to investigate neutral \ion{C}{i} lines due to its extremely carbon enhancement. J0815 + 4729 is also rich in oxygen, making it possible to analyze the IR \ion{O}{i} triplet. 

We perform an independent analysis of three indicators of the carbon abundance, that is, the CH and C$_2$ molecular lines and the \ion{C}{i} atomic lines under the assumption of 1D LTE. At the same time, for \ion{C}{i}, a detailed NLTE analysis of carbon was carried out with a newly constructed atomic model, and the NLTE abundance of oxygen from the \ion{O}{i} infrared triplet was also investigated.

\section{Data}

\subsection{Observed spectra}
For J0815+4729, high-resolution spectra were obtained with high resolution echelle spectrograph (HIRES) \citep{Vogt1994} on the Keck I telescope at the Maunakea observatory. The wavelength range spanned $\lambda$340-780\,nm with a resolving power of R $\sim$ 37\,500. Seven exposures of 2400s each were acquired, and the individual reduced spectra yielded signal-to-noise ratios of 15, 21, 31, 40, and 42 at $\lambda =$ 395, 440, 518, 670, and 775 nm, respectively. The coadded spectrum adopted here was RV-corrected, normalized, and rebinned into three regions, the blue (340-473~nm), green (478-631~nm), and red (632-782~nm). We refer to \citet{Gonzalez2020} for more details about the observations and data reduction.

\subsection{Model atom}
Our carbon and oxygen model atoms were newly constructed following the same method as described in \citet{Osorio2015,Osorio2019,Osorio2020}. The carbon model atom had 84 neutral, 48 singly ionized, and the ground state of the doubly ionized levels, and the oxygen model atom had 74 neutral, 61 singly ionized, and the ground state of the doubly ionized levels. A detailed description of the model atoms will be presented in a forthcoming paper (Zhang et al. in prep.).

\subsection{Line data}
The line data for neutral carbon and oxygen transitions were obtained from NIST\footnote{\url{https://physics.nist.gov/PhysRefData/ASD/lines_form.html}}. We included five \ion{C}{i} lines in our analysis. For oxygen, only the \ion{O}{i} $\lambda$777\,nm triplet line was strong enough to be visible within our spectral range.  The selected atomic lines are listed in Table~\ref{tab:linelist}. The molecular line data were adopted from Kurucz’s compiled line lists\footnote{\url{http://kurucz.harvard.edu/}} \citep{Kurucz2011,Kurucz2018}. We employed the CH $G$-band lines $\lambda$426 $-$ 432~nm and two sets of CH lines adjacent to the CN band at 388.3 nm. The adopted C$_{\rm 2}$ molecular lines were at $\lambda$516.5\,nm.

\begin{table*}[ht!]
\caption{Line information and abundances of the \ion{C}{i} and \ion{O}{i} lines \newline}
\label{tab:linelist}
\centering
\begin{tabular}{clclccc}
\hline\hline
 Lines (nm) & Transitions & $\chi$ (eV) & log~$gf$ & A(X)$_{\rm LTE}$ & A(X)$_{\rm NLTE}$ & $\Delta_{\rm NLTE}$\\
\hline
\multicolumn{7}{c}{\ion{C}{i}}\\
\hline
 505.2144 & 3$s$ $^{1}\rm{P}^{\rm{o}}_{\rm{1}}$ $-$ 4$p ^{1}_{}\rm{D}_{{\rm 2}}$ & 7.685 & $-$1.303$^a$ & 8.16 & 8.08 & $-$0.08\\
 538.0325 & 3$s$ $^{1}\rm{P}^{\rm{o}}_{\rm{1}}$ $-$ 4$p ^{1}\rm{P}_{\rm{1}}$ & 7.685 & $-$1.616$^a$ & 8.09 & 8.01 & $-$0.08\\
 658.7620 & 3$p$ $^{1}\rm{P}_{\rm{1}}$ $-$ 4$d ^{1}\rm{P}^{\rm{o}}_{\rm{1}}$ & 8.537 & $-$1.003$^{b,a}$ & 8.15 & 8.11 & $-$0.04\\
 711.1461 & 3$p$ $^{3}\rm{D}_{\rm{1}}$ $-$ 4$d ^{3}\rm{F}^{\rm{o}}_{{\rm 2}}$ & 8.640 & $-$1.085$^a$ & 8.15 & 8.11 & $-$0.04\\
 711.3172 & 3$p$ $^{3}\rm{D}_{\rm{3}}$ $-$ 4$d ^{3}\rm{F}^{\rm{o}}_{\rm{4}}$ & 8.647 & $-$0.773$^a$ & 8.13 & 8.08 & $-$0.05\\
\hline
\multicolumn{7}{c}{\ion{O}{i}}\\
\hline
 777.1944 & 3$s$ $^{5}\rm{S}^{\rm{o}}_{{\rm 2}}$ $-$ 3$p ^{5}\rm{P}_{\rm 3}$ &  9.146   &  0.369$^c$ & 7.36 & 7.25 & $-$0.11\\
 777.4166 & 3$s$ $^{5}\rm{S}^{\rm{o}}_{{\rm 2}}$ $-$ 3$p ^{5}\rm{P}_{\rm 2}$ &  9.146   &  0.223$^c$ & 7.36 & 7.25 & $-$0.11\\
 777.5388 & 3$s$ $^{5}\rm{S}^{\rm{o}}_{{\rm 2}}$ $-$ 3$p ^{5}\rm{P}_{\rm{1}}$ &  9.146  &   0.002$^c$ & 7.36 & 7.25 & $-$0.11\\
\hline
\end{tabular}
\tablefoot{The wavelengths are in air.
References to the log\,$gf$ values are a: \citet{Hibbert1993}, b: \citet{Luo1989}, and c: \citet{Hibbert1991}. $\Delta_\mathrm{NLTE}$ denotes the NLTE abundance correction, defined as $\Delta_\mathrm{NLTE} = \log\epsilon_\mathrm{NLTE} - \log\epsilon_\mathrm{LTE}$.}
\end{table*}

\section{Method}
The model atmospheres were newly computed using Sbordone's port \citep{Sbordone2007} of ATLAS9\footnote{\url{https://github.com/callendeprieto/mkk-atlas9}} \citep{Kurucz2005}. The departure coefficients were calculated using the code TLUSTY \citep{Hubeny2021}. All synthetic LTE and NLTE spectra were computed with Synple\footnote{\url{https://github.com/callendeprieto/synple}}, which is a Python interface to the spectral synthesis code SYNSPEC \citep{Hubeny2021} and also includes a set of related utilities.

We carried out the abundance measurements using two independent fitting strategies: a line-by-line best-fit method for atomic transitions, and a $\chi^2$ minimization of synthetic molecular band profiles against the observed spectra. When we fit individual neutral atomic lines, the radial-tangential macroturbulent velocity $v_{\rm mac}$ and carbon abundance 
A(C) were treated as free parameters to achieve optimal agreement with observations. In contrast, the global $\chi^2$ minimization technique adopts a single uniform 
$v_{\rm mac}$ and single A(C) value to simultaneously fit all molecular bands. We adopted solar abundances of log\,$\epsilon_{\rm C}$ = 8.46 and log\,$\epsilon_{\rm O}$ = 8.69 \citep{Asplund2021} for reference.

\section{J0815+4729}\label{sec:j0815}
\subsection{Stellar parameters}
We independently determined the stellar parameters: we determined the effective temperature and surface gravity by best-fitting the \ion{H}{i} Balmer lines separately in an iterative way, and we derived [Fe/H] by analyzing three \ion{Fe}{i} lines ($\lambda$371.993\,nm, $\lambda$382.043\,nm, and $\lambda$385.991\,nm). We obtained the same temperature and surface gravity as \citet{Gonzalez2020}, that is, $T_{\rm eff} =$ 6050 $\pm$ 100\,K and log\,$g = 4.6$ $\pm$ 0.2\, dex. We derived an LTE iron abundance of 2.05 $\pm$ 0.11 dex, which agrees well with 2.01 $\pm$ 0.14 dex from \citet{Gonzalez2020}.

Because we have not yet established a proved and coherent iron model atom following the same approach as our carbon and oxygen model atoms, we adopted the NLTE correction grids from \citet{Kovalev2018} for the derived \ion{Fe}{i} abundance at the current stage. This yielded a mean NLTE correction of $\sim$ +0.2~dex. We found that an increase of 0.2\,dex in [Fe/H] affects the abundances of C and O in J0815+4729 very little. Therefore, we adopted the LTE [Fe/H] value of $-$5.49 for the subsequent analysis. The microturbulence was adopted as $\xi_t =$ 1.5 km s$^{-1}$, following the empirical scale for metal-poor dwarf stars \citep{Barklem2005, Gonzalez2020}. The final stellar parameters $T_{\rm eff}$/log\,$g$/[Fe]/$\xi_t$ of J0815+4729 are 6050~K/4.6~dex/$-$5.49~dex/1.5~km~s$^{-1}$. 

\subsection{\ion{C}{i}, CH, and C$_2$}

\begin{figure*}
\sidecaption
\raisebox{1.4cm}{
\begin{minipage}{12cm}
\centering
\includegraphics[width=0.49\linewidth]{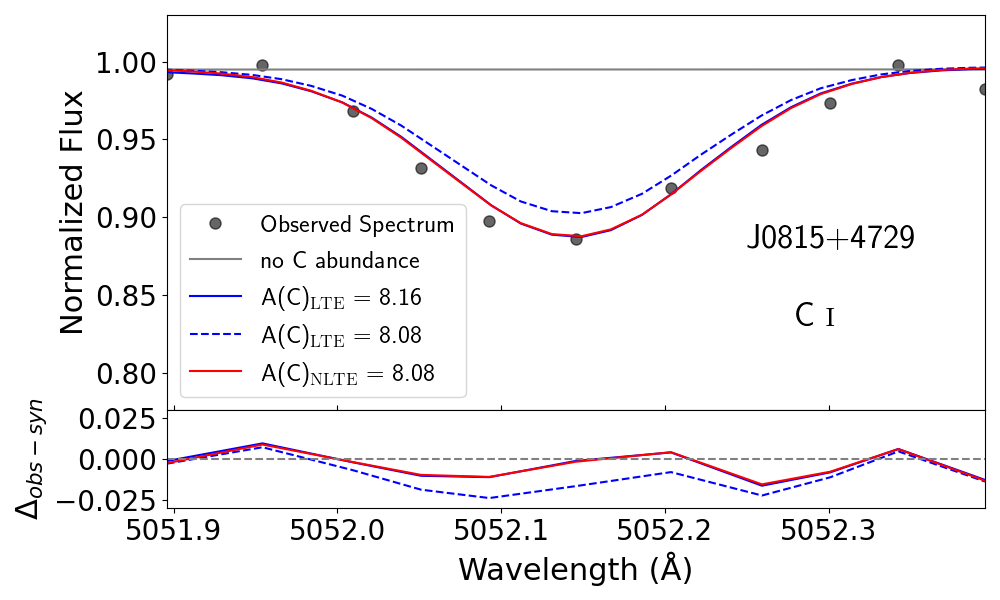}
\hfill
\includegraphics[width=0.49\linewidth]{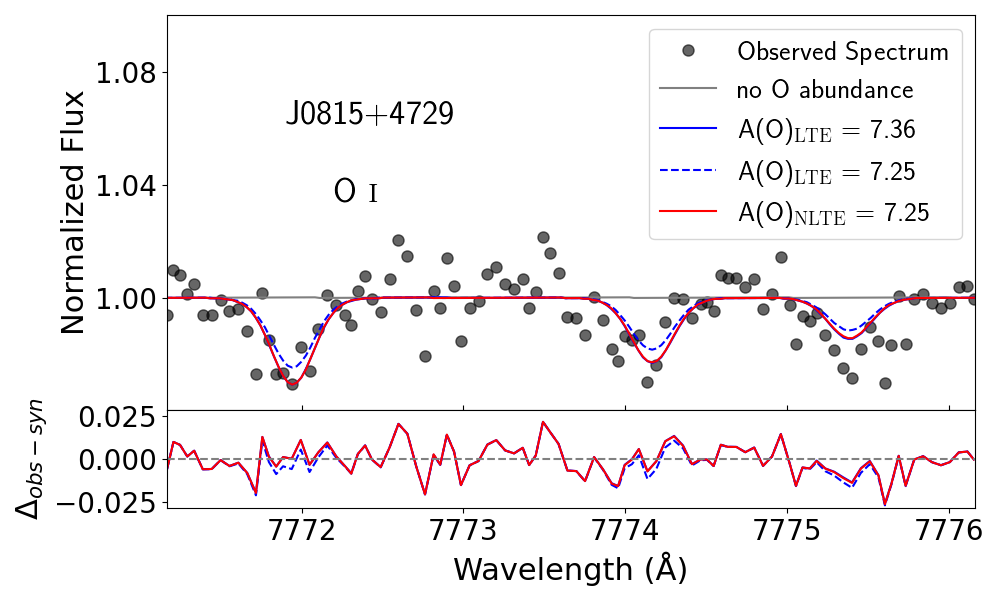}
\vspace{0.3cm}
\includegraphics[width=0.49\linewidth]{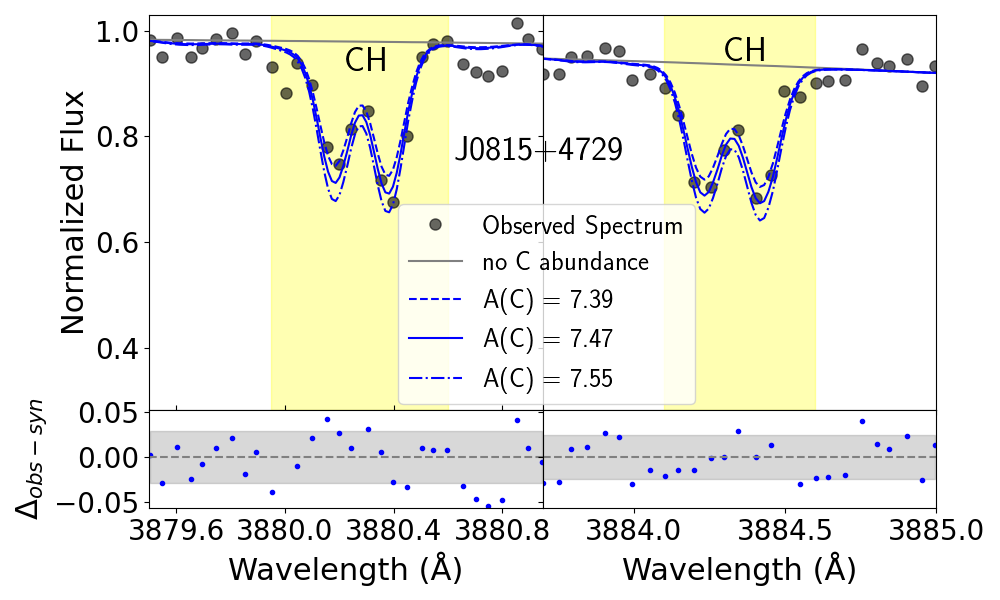}
\hfill
\includegraphics[width=0.49\linewidth]{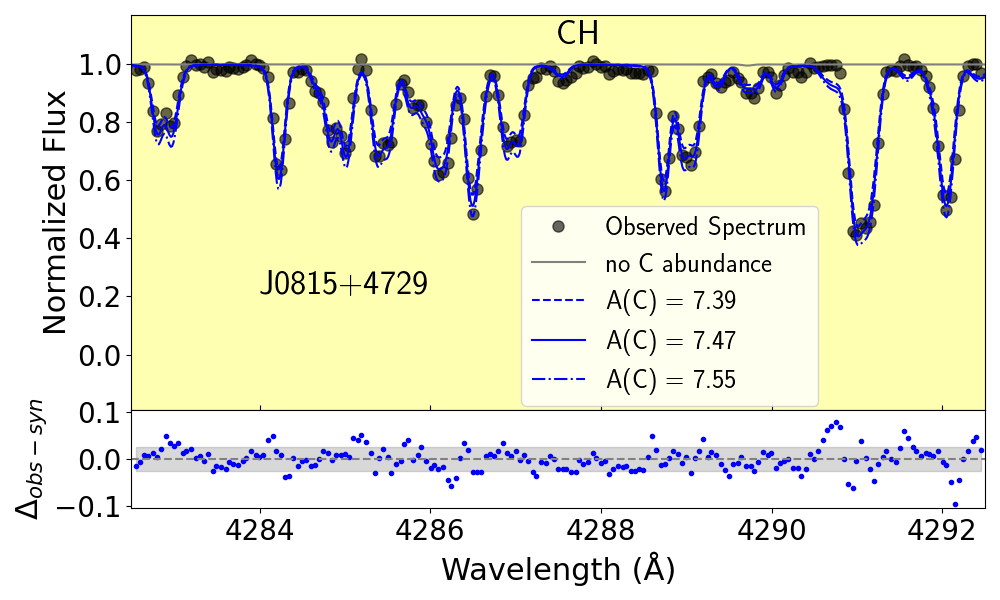}
\vspace{0.3cm}
\includegraphics[width=0.49\linewidth]{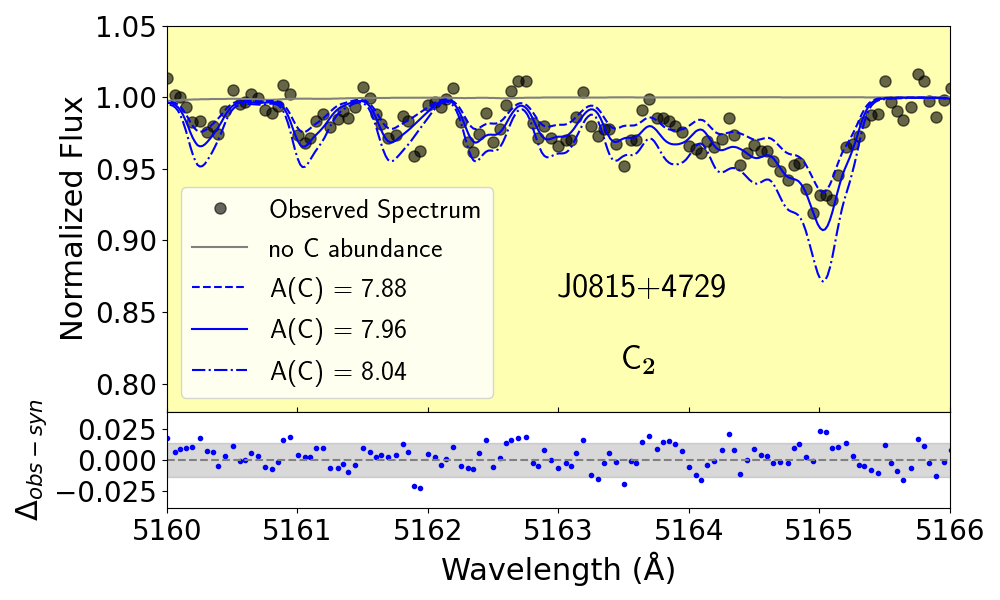}
\end{minipage}
}
\caption{Upper two panels: NLTE (solid red line) and LTE (solid blue line) synthetic spectra compared with the observed spectra for \ion{C}{i} $\lambda$505.2144\,nm (upper left) and the \ion{O}{i} triplet (upper right) in J0815+4729. The dashed blue line indicates the LTE synthetic spectrum with the same abundance as the best-fitting NLTE value. The solid gray line indicates the synthesized spectrum without C or O abundances. Middle two panels: Best-fitting results (solid blue line) for two CH indicators in J0815+4729 under LTE, and the dashed and dash-dotted lines represent abundance shifts of $-$0.08 and +0.08 dex, respectively. Bottom panel: Best fit for the C$_2$ lines. The evaluation regions for the carbon molecular features when determining carbon abundances adopting the $\chi^2$ method are indicated in yellow. The lower part of each panel displays the residuals ($\Delta_{obs-syn}$).}
\label{fig:metal-poor_bestfitting}
\end{figure*}

The carbon abundances derived from the \ion{C}{i} lines in J0815+4729 are 8.14 $\pm$ 0.07 ([C/Fe] = 5.17 $\pm$ 0.07) and 8.08 $\pm$ 0.07 ([C/Fe] = 5.11 $\pm$ 0.07) under LTE and NLTE, respectively. We also independently investigated the selected molecular CH lines and the C$_2$ Swan band and obtained carbon abundances of 7.47 $\pm$ 0.17 ([C/Fe] = 4.50 $\pm$ 0.17) and 7.96 $\pm$ 0.08 ([C/Fe] = 4.99 $\pm$ 0.08) in LTE, respectively. 

The best fits to selected spectral features of J0815+472 are presented in Figure~\ref{fig:metal-poor_bestfitting}. The upper left panel presents an example of LTE and NLTE profiles for the \ion{C}{i} $\lambda$505.2~nm line, and its NLTE correction is $-$0.08. The lower panels show the examples of the fittings for CH and C$_2$. 

\begin{table}[ht!]
\centering
\caption{C and O abundances inferred for J0815+4729}
\label{tab:sum_abu}
\begin{tabular}{lcccc}
\hline\hline 
Indicator & & LTE & NLTE & $\Delta_\mathrm{NLTE}$\\
\hline
$A(\mathrm{C})_{\ion{C}{I}}$ & & $8.14\pm0.07$ & $8.08\pm0.07$ & $-0.06$ \\
$A(\mathrm{C})_\mathrm{CH\;\text{G-band}}$ & & $7.47\pm0.17$ & & \\
$A(\mathrm{C})_\mathrm{CH\;\text{near\;CN}}$ & & $7.47\pm0.17$ & & \\
$A(\mathrm{C})_{\mathrm{C}_2}$ & & $7.96\pm0.08$ & & \\
$A(\mathrm{O})_{\ion{O}{I}}$ & & 
$7.36\pm0.11$ & $7.25\pm0.11$ & $-0.11$ \\
\hline
\end{tabular}
\end{table}

 The derived abundances and NLTE corrections for individual atomic lines are listed in Table~\ref{tab:linelist}. The carbon abundances derived from different tracers are presented in Table\,\ref{tab:sum_abu}. The abundance uncertainties were determined by taking the square root of the sum of all individual uncertainties in quadrature, including the propagated errors in the atmospheric parameters, $\delta T_{\rm eff} = 100$\,K, $\delta\,{\rm log}\,g = 0.20$\,dex, $\delta \xi_t = 0.5 \,{\rm km\,s}^{-1}$, and the dispersion from line-to-line (for the atomic lines) or statistical error (for the molecular lines). Since the neutral carbon lines are weaker in metal-poor stars and usually subject to smaller 3D effects than CH and C$_2$ \citep{Caffau2010, Dobrovolskas2013, Alexeeva2015}, we favor the carbon abundance determined from the 1D NLTE \ion{C}{i} analysis, that is, A(C) = 8.08$\pm$0.07 for J0815+4715.

Our result for the CH lines is consistent with the value reported by \citet{Gonzalez2020}, 7.43 $\pm$ 0.17. Similarly, our C$_2$-based abundance agrees with their C$_2$-based value of 8.04 $\pm$ 0.08 within the error bars. 

\subsection{Doubts on the reliability of CH}
We found significantly higher abundances from C$_2$ than from CH, and the difference is 0.5$-$0.6 dex. A discrepancy like this has been reported in previous studies of carbon-enhanced metal-poor stars \citep{Christlieb2004, Hill2000, Behara2010}. 

For J0815+4729, a similar discrepancy, $\sim$\,0.6 dex, is also apparent between the abundances inferred from the \ion{C}{i} and CH lines. This effect was reported in previous studies \citep{Tomkin1992, Behara2010}. At the same time, our NLTE \ion{C}{i} abundance (8.08 $\pm$ 0.07) is much closer to the C$_2$-based abundance (7.96 $\pm$ 0.08). 

Compared to C$_2$ lines, \citet{Christlieb2004} demonstrated that the CH lines are more sensitive to the temperature (see their Table~5). When they increased the effective temperature from 5100\,K to 5500\,K, the abundance discrepancy between CH and C$_2$ was removed. However, such a high temperature is not acceptable for the star they studied.
\cite{Alexeeva2015} examined the differences of carbon between \ion{C}{i} and CH in four stars in common with \cite{Tomkin1992} and found that the discrepancy decreased from $+$0.4 to $-$0.02 dex when they increased the effective temperature by 130 K.  

Our tests indicate that increasing $T_{\rm eff}$ by 200 K weakens the synthetic carbon molecular lines, which in turn leads to a higher inferred carbon abundance from CH and C$_2$, specifically, 7.77 (increased by 0.3) and 8.17 (increased by 0.21). In contrast, the synthetic \ion{C}{i} lines are strengthened, corresponding to a lower carbon abundance of 7.95 (reduced by 0.13). It is clear that an increase in the temperature alone cannot eliminate the discordance among the three indicators because it would result in an abnormally high abundance of C$_2$. 

The 3D effects on the abundance determination from CH lines in red giants was evaluated by \citet{Collet2007}. The 3D $-$ 1D abundance corrections are negative overall and can reach $-$0.8 with stellar parameters of 5500\,K, ${\rm log}\,g = 2.2$\,dex, and [Fe/H] $=$ $-$3.0\,dex. \citet{Eitner2025} also conducted a detailed study of 3D stellar atmosphere models for CEMP stars and concluded that the 3D effects are always negative for CH and amount to approximately $-$0.6~dex for a model with $T_{\rm eff}$ $=$ 5750\,K, log $g$ $=$ 4.5, and [Fe/H] $=$ $-$5.5. This implies that the current 3D models do not assist in reducing the differences among the three carbon indicators and would further enlarge them.

The NLTE effects are generally prominent in giant stars and in metal-poor stars because these objects typically have low number densities of particles, making them prone to deviations from LTE. According to \citet{Popa2023}, the NLTE corrections for molecular CH lines can range from +0.04 for the Sun to +0.21 dex for a red giant with [Fe/H] = $-$4.0. Although J0815+4729 is a dwarf star, the discrepancies between the \ion{C}{i}/C$_2$ and CH spectral features might partially arise from NLTE effects of molecular features, owing to its extremely metal-poor character.

\cite{Santos2025} derived [C/Fe] ratios for ~200 stars (–2.5 < [Fe/H] < +0.5) from two independent synthetic spectral grids using CH molecular bands in the G band. They found that carbon abundances derived from the CH G-band lines strongly relied on the modeling particulars. Even minor intrinsic differences between synthetic models for the crowded, blended 430\,nm CH region can produce large offsets up to 0.8 dex in [C/Fe]. This further casts doubt on the reliability of the CH band as a carbon abundance tracer.

\subsection{\ion{O}{i}}

We analyzed the \ion{O}{i} $\lambda$777 nm triplet in J0815+4729 and derived abundances of 7.36 $\pm 0.11$ ([O/Fe] $=$ 4.16 $\pm 0.11$, LTE) and 7.25 $\pm 0.11$ ([O/Fe] $=$ 4.05 $\pm 0.11$, NLTE), with an NLTE correction of $-$0.11 dex, as shown in Table\,\ref{tab:sum_abu} (see the fitting result in the upper right panel of Figure~\ref{fig:metal-poor_bestfitting}). Our LTE result is consistent with the LTE value reported by \citet{Gonzalez2020} (7.23 $\pm$ 0.14). We derived a carbon-to-oxygen ratio of the star of [C/O] $=$ 1.06 $\pm$ 0.11.

\section{Conclusions}
The difference in carbon abundances derived from \ion{C}{i}/C$_2$ lines and CH is approximately 0.6\,dex, which cannot be explained by 3D effects. NLTE effects for molecular lines might offer a plausible explanation, but further investigation is necessary. Part of the discrepancy might be caused by systematic errors in the stellar parameters, especially the effective temperature. According to a previous study, owing to severe line blending, the crowded G-band CH region is subject to substantial model-dependent systematic uncertainties.

Thus, we highlighted the risk of using CH lines as effective tracers to understand the chemical enrichment history of the cosmos. According to our result, we recommend using \ion{C}{i} transitions instead of CH lines in these hyper-iron-poor stars when reliable \ion{C}{i} lines are available.
 
\begin{figure}
    \centering
    \includegraphics[width=0.8\linewidth]{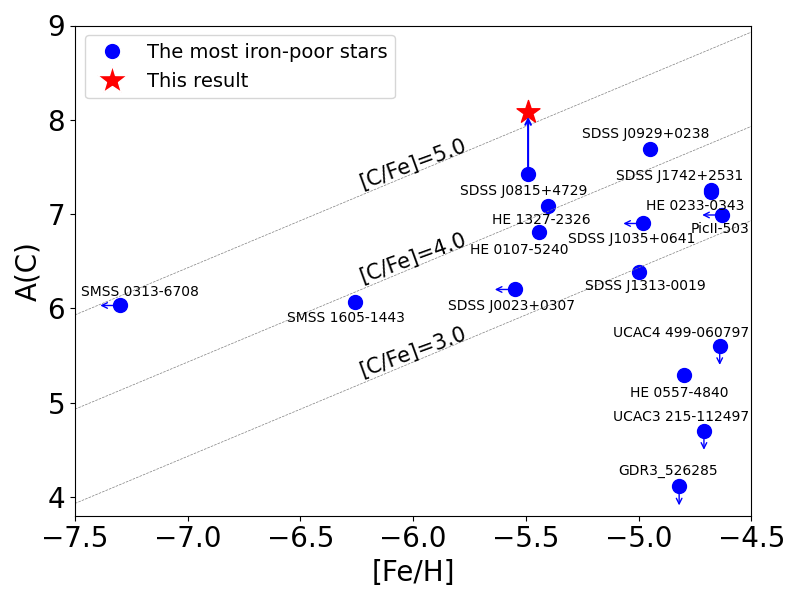}
    \includegraphics[width=0.88\linewidth]{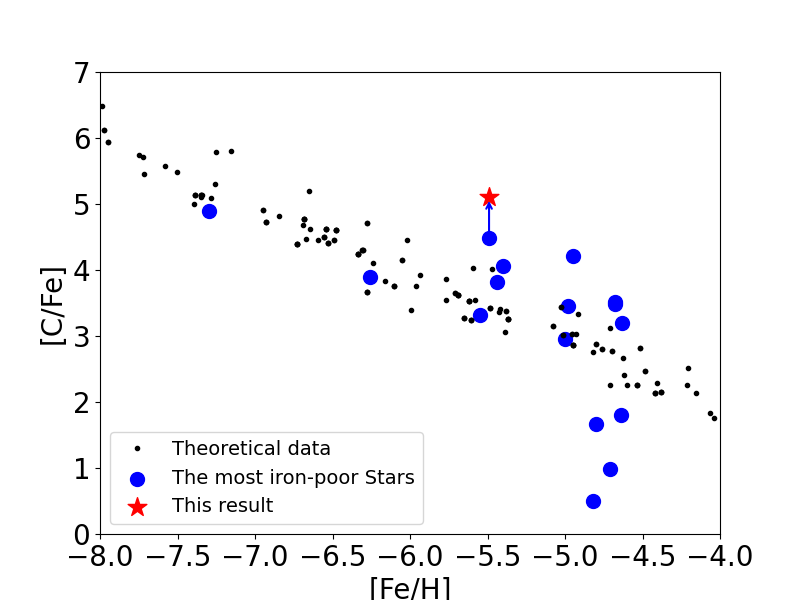}
    \caption{Top panel: Carbon abundance vs. iron abundance for the 16 known most iron-poor stars of [Fe/H] $<$ $-$4.5. 
The red star represents our carbon result. Bottom panel: [C/Fe] ratios as a function of [Fe/H], predicted by the supernova models (black dots; with iron multiplied by an arbitrary factor) and the abundances found in the most iron-poor stars known (blue circles).}
    \label{fig:metal-poor_C}
\end{figure}

The distribution of 16 stars with the lowest iron abundances ([Fe/H] $<$ $-$4.5) in the A(C) versus [Fe/H] plane is displayed in the upper panel of Figure~\ref{fig:metal-poor_C}. The star J0815+4729 has the highest carbon content at [Fe/H] $<$ $-$4.5. 

\citet{Gonzalez2020} derived LTE upper limits of [Sr/H]$< -$4.47 and [Ba/H] $< -$3.58, corresponding to [Sr/Fe] < 1.02 and [Ba/Fe] < 1.91 with [Fe/H]$= -$5.49 for J0815+4729. They placed the star (A(C) $=$ 7.43) in the upper limit of the low-carbon band in the A(C) $-$[Fe/H] diagram, which suggests that it is a CEMP-no star \citep{Bonifacio2015}. However, these upper limits for [Sr/Fe] and [Ba/Fe] are still relatively weak constraints and cannot definitively exclude mild AGB pollution. In addition, our updated abundance calculations move the star’s position toward the high-carbon band in the A(C) $-$ [Fe/H] plane. This location suggests that J0815+4729 might have experienced pollution from an AGB companion. Further observations searching for RV variations are underway to test this scenario.

In the lower panel of Figure \ref{fig:metal-poor_C}, we plot the theoretically predicted [C/Fe] values along with [Fe/H], represented by small black dots. These values were derived from the yields of the related elements calculated by \cite{Heger2010} and involve the conversion from mass fraction to particle fraction, following the same approach as described in \cite{Allende2015}. The 16 most iron-poor stars are overlaid on the plot. 

Four stars agree with the theoretical trend, suggesting that they might have formed from gas that was enriched by supernovae resulting from metal-free stars in the mass range 10$-$100~$M_{\odot}$. However, the remaining 12 iron-poor stars, including J0815+4729, fit this framework only poorly, indicating that their chemical composition is linked to a different type of source and/or that the models we considered do not cover a sufficiently broad range of explosion energies. 
Models with lower explosion energies combined with significant fallback of iron-rich material onto the compact remnant are expected to favor the observations, since higher explosion energies expel most iron, resulting in lower [C/Fe]. Suitable progenitors lie around 20–30 M$_\sun$, together with moderate internal mixing.
The distinctive high carbon abundance of J0815+4729 may provide insights into the early chemical evolution of the Galaxy and the nature of the first stars. 

Our study confirms that J0815+4729 is a significantly oxygen-enhanced star with A(O)$_{\rm NLTE}$ = 7.25 $\pm$ 0.11, as previously found by \citet{Gonzalez2020}. The carbon-to-oxygen ratio of the star is [C/O] = 1.06 ± 0.11.

\begin{acknowledgements}
J.B.Z. gratefully acknowledges support from National Key R\&D Program of China, Grant No. 2023YFF0714800, the Scientific Instrument Developing Project of the Chinese Academy of Sciences, Grant No. ZDKYYQ20220009, the China Scholarship Council, International partnership program of the Chinese academy of sciences, Grant No. 178GJHZ2022047GC. C.A.P., J,G.H., R.R, and D.A. acknowledge financial support from the Spanish Ministry of Science, Innovation and Universities (MICIU), under grants PID2023-149982NB-I00 and PID2023-146453NB-I00, and from the Agencia Estatal de Investigación (AEI) through the Severo Ochoa Centre of Excellence accreditation awarded to the Instituto de Astrofísica de Canarias, grant CEX2025-001609-S, funded by MICIU/AEI/10.13039/501100011033. J.R.S. acknowledges National Natural Science Foundation of China under Grant Nos. 12090040/4, 12427804. Some of the data presented herein were obtained at Keck Observatory, which is a private 501(c)3 non-profit organization operated as a scientific partnership among the California Institute of Technology, the University of California, and the National Aeronautics and Space Administration. The Observatory was made possible by the generous financial support of the W. M. Keck Foundation. 
\end{acknowledgements}

\bibliographystyle{aa} 
\bibliography{aa61833-26_proofs.bib}

\begin{thebibliography}{50}
\expandafter\ifx\csname natexlab\endcsname\relax\def\natexlab#1{#1}\fi

\bibitem[{{Aguado} {et~al.}(2018{\natexlab{a}}){Aguado}, {Allende Prieto},
  {Gonz{\'a}lez Hern{\'a}ndez}, \& {Rebolo}}]{Aguado2018a}
{Aguado}, D.~S., {Allende Prieto}, C., {Gonz{\'a}lez Hern{\'a}ndez}, J.~I., \&
  {Rebolo}, R. 2018{\natexlab{a}}, \apjl, 854, L34

\bibitem[{{Aguado} {et~al.}(2018{\natexlab{b}}){Aguado}, {Gonz{\'a}lez
  Hern{\'a}ndez}, {Allende Prieto}, \& {Rebolo}}]{Aguado2018b}
{Aguado}, D.~S., {Gonz{\'a}lez Hern{\'a}ndez}, J.~I., {Allende Prieto}, C., \&
  {Rebolo}, R. 2018{\natexlab{b}}, \apjl, 852, L20

\bibitem[{{Aguado} {et~al.}(2019){Aguado}, {Gonz{\'a}lez Hern{\'a}ndez},
  {Allende Prieto}, \& {Rebolo}}]{Aguado2019}
{Aguado}, D.~S., {Gonz{\'a}lez Hern{\'a}ndez}, J.~I., {Allende Prieto}, C., \&
  {Rebolo}, R. 2019, \apjl, 874, L21

\bibitem[{{Alexeeva} \& {Mashonkina}(2015)}]{Alexeeva2015}
{Alexeeva}, S.~A. \& {Mashonkina}, L.~I. 2015, \mnras, 453, 1619

\bibitem[{{Allende Prieto} {et~al.}(2015){Allende Prieto},
  {Fern{\'a}ndez-Alvar}, {Aguado}, {Gonz{\'a}lez Hern{\'a}ndez}, {Rebolo},
  {Lee}, {Beers}, {Rockosi}, \& {Ge}}]{Allende2015}
{Allende Prieto}, C., {Fern{\'a}ndez-Alvar}, E., {Aguado}, D.~S., {et~al.}
  2015, \aap, 579, A98

\bibitem[{{Aoki} {et~al.}(2007){Aoki}, {Beers}, {Christlieb}, {Norris}, {Ryan},
  \& {Tsangarides}}]{Aoki2007}
{Aoki}, W., {Beers}, T.~C., {Christlieb}, N., {et~al.} 2007, \apj, 655, 492

\bibitem[{{Asplund} {et~al.}(2021){Asplund}, {Amarsi}, \&
  {Grevesse}}]{Asplund2021}
{Asplund}, M., {Amarsi}, A.~M., \& {Grevesse}, N. 2021, \aap, 653, A141

\bibitem[{{Barklem} {et~al.}(2005){Barklem}, {Christlieb}, {Beers}, {Hill},
  {Bessell}, {Holmberg}, {Marsteller}, {Rossi}, {Zickgraf}, \&
  {Reimers}}]{Barklem2005}
{Barklem}, P.~S., {Christlieb}, N., {Beers}, T.~C., {et~al.} 2005, \aap, 439,
  129

\bibitem[{Beers \& Christlieb(2005)}]{Beers2005}
Beers, T.~C. \& Christlieb, N. 2005, ARA\&A, 43, 531

\bibitem[{{Behara} {et~al.}(2010){Behara}, {Bonifacio}, {Ludwig}, {Sbordone},
  {Gonz{\'a}lez Hern{\'a}ndez}, \& {Caffau}}]{Behara2010}
{Behara}, N.~T., {Bonifacio}, P., {Ludwig}, H.-G., {et~al.} 2010, \aap, 513,
  A72

\bibitem[{{Bonifacio} {et~al.}(2025){Bonifacio}, {Caffau}, {Fran{\c{c}}ois}, \&
  {Spite}}]{Bonifacio2025}
{Bonifacio}, P., {Caffau}, E., {Fran{\c{c}}ois}, P., \& {Spite}, M. 2025,
  \aapr, 33, 2

\bibitem[{{Bonifacio} {et~al.}(2015){Bonifacio}, {Caffau}, {Spite}, {Limongi},
  {Chieffi}, {Klessen}, {Fran{\c{c}}ois}, {Molaro}, {Ludwig}, {Zaggia},
  {Spite}, {Plez}, {Cayrel}, {Christlieb}, {Clark}, {Glover}, {Hammer}, {Koch},
  {Monaco}, {Sbordone}, \& {Steffen}}]{Bonifacio2015}
{Bonifacio}, P., {Caffau}, E., {Spite}, M., {et~al.} 2015, \aap, 579, A28

\bibitem[{{Bonifacio} {et~al.}(2018){Bonifacio}, {Caffau}, {Spite}, {Spite},
  {Sbordone}, {Monaco}, {Fran{\c{c}}ois}, {Plez}, {Molaro}, {Gallagher},
  {Cayrel}, {Christlieb}, {Klessen}, {Koch}, {Ludwig}, {Steffen}, {Zaggia}, \&
  {Abate}}]{Bonifacio2018}
{Bonifacio}, P., {Caffau}, E., {Spite}, M., {et~al.} 2018, \aap, 612, A65

\bibitem[{{Caffau} {et~al.}(2011){Caffau}, {Bonifacio}, {Fran{\c{c}}ois},
  {Sbordone}, {Monaco}, {Spite}, {Spite}, {Ludwig}, {Cayrel}, {Zaggia},
  {Hammer}, {Randich}, {Molaro}, \& {Hill}}]{Caffau2011}
{Caffau}, E., {Bonifacio}, P., {Fran{\c{c}}ois}, P., {et~al.} 2011, \nat, 477,
  67

\bibitem[{{Caffau} {et~al.}(2016){Caffau}, {Bonifacio}, {Spite}, {Spite},
  {Monaco}, {Sbordone}, {Fran{\c{c}}ois}, {Gallagher}, {Plez}, {Zaggia},
  {Ludwig}, {Cayrel}, {Koch}, {Steffen}, {Salvadori}, {Klessen}, {Glover}, \&
  {Christlieb}}]{Caffau2016}
{Caffau}, E., {Bonifacio}, P., {Spite}, M., {et~al.} 2016, \aap, 595, L6

\bibitem[{{Caffau} {et~al.}(2010){Caffau}, {Ludwig}, {Bonifacio}, {Faraggiana},
  {Steffen}, {Freytag}, {Kamp}, \& {Ayres}}]{Caffau2010}
{Caffau}, E., {Ludwig}, H.-G., {Bonifacio}, P., {et~al.} 2010, \aap, 514, A92

\bibitem[{{Chiti} {et~al.}(2026){Chiti}, {Placco}, {Pace}, {Ji}, {Prabhu},
  {Cerny}, {Limberg}, {Stringfellow}, {Drlica-Wagner}, {Atzberger}, {Choi},
  {Crnojevi{\'c}}, {Ferguson}, {Kallivayalil}, {No{\"e}l}, {Riley}, {Sand},
  {Simon}, {Walker}, {Bom}, {Carballo-Bello}, {James},
  {Mart{\'\i}nez-V{\'a}zquez}, {Medina}, \& {Vivas}}]{Chiti2026}
{Chiti}, A., {Placco}, V.~M., {Pace}, A.~B., {et~al.} 2026, Nature Astronomy,
  10, 830

\bibitem[{{Christlieb} {et~al.}(2002){Christlieb}, {Bessell}, {Beers},
  {Gustafsson}, {Korn}, {Barklem}, {Karlsson}, {Mizuno-Wiedner}, \&
  {Rossi}}]{Christlieb2002}
{Christlieb}, N., {Bessell}, M.~S., {Beers}, T.~C., {et~al.} 2002, \nat, 419,
  904

\bibitem[{{Christlieb} {et~al.}(2004){Christlieb}, {Gustafsson}, {Korn},
  {Barklem}, {Beers}, {Bessell}, {Karlsson}, \&
  {Mizuno-Wiedner}}]{Christlieb2004}
{Christlieb}, N., {Gustafsson}, B., {Korn}, A.~J., {et~al.} 2004, \apj, 603,
  708

\bibitem[{{Collet} {et~al.}(2007){Collet}, {Asplund}, \&
  {Trampedach}}]{Collet2007}
{Collet}, R., {Asplund}, M., \& {Trampedach}, R. 2007, \aap, 469, 687

\bibitem[{{Dobrovolskas} {et~al.}(2013){Dobrovolskas}, {Ku{\v{c}}inskas},
  {Steffen}, {Ludwig}, {Prakapavi{\v{c}}ius}, {Klevas}, {Caffau}, \&
  {Bonifacio}}]{Dobrovolskas2013}
{Dobrovolskas}, V., {Ku{\v{c}}inskas}, A., {Steffen}, M., {et~al.} 2013, \aap,
  559, A102

\bibitem[{{Eitner} {et~al.}(2025){Eitner}, {Bergemann}, {Hoppe}, {Storm},
  {Lipatova}, {Glover}, {Klessen}, {Nordlund}, \& {Popovas}}]{Eitner2025}
{Eitner}, P., {Bergemann}, M., {Hoppe}, R., {et~al.} 2025, \aap, 703, A199

\bibitem[{{Frebel} {et~al.}(2005){Frebel}, {Aoki}, {Christlieb}, {Ando},
  {Asplund}, {Barklem}, {Beers}, {Eriksson}, {Fechner}, {Fujimoto}, {Honda},
  {Kajino}, {Minezaki}, {Nomoto}, {Norris}, {Ryan}, {Takada-Hidai},
  {Tsangarides}, \& {Yoshii}}]{Frebel2005}
{Frebel}, A., {Aoki}, W., {Christlieb}, N., {et~al.} 2005, \nat, 434, 871

\bibitem[{{Gonz{\'a}lez Hern{\'a}ndez} {et~al.}(2023){Gonz{\'a}lez
  Hern{\'a}ndez}, {Aguado}, {Allende-Prieto}, {Burgasser}, \&
  {Rebolo}}]{Gonzalez2023}
{Gonz{\'a}lez Hern{\'a}ndez}, J.~I., {Aguado}, D.~S., {Allende-Prieto}, C.,
  {Burgasser}, A., \& {Rebolo}, R. 2023, in Memorie della Societa Astronomica
  Italiana, Vol.~94, 77

\bibitem[{{Gonz{\'a}lez Hern{\'a}ndez} {et~al.}(2020){Gonz{\'a}lez
  Hern{\'a}ndez}, {Aguado}, {Allende Prieto}, {Burgasser}, \&
  {Rebolo}}]{Gonzalez2020}
{Gonz{\'a}lez Hern{\'a}ndez}, J.~I., {Aguado}, D.~S., {Allende Prieto}, C.,
  {Burgasser}, A.~J., \& {Rebolo}, R. 2020, \apjl, 889, L13

\bibitem[{{Hansen} {et~al.}(2014){Hansen}, {Hansen}, {Christlieb}, {Yong},
  {Bessell}, {Garc{\'\i}a P{\'e}rez}, {Beers}, {Placco}, {Frebel}, {Norris}, \&
  {Asplund}}]{Hansen2014}
{Hansen}, T., {Hansen}, C.~J., {Christlieb}, N., {et~al.} 2014, \apj, 787, 162

\bibitem[{{Heger} \& {Woosley}(2010)}]{Heger2010}
{Heger}, A. \& {Woosley}, S.~E. 2010, \apj, 724, 341

\bibitem[{{Hibbert} {et~al.}(1991){Hibbert}, {Biemont}, {Godefroid}, \&
  {Vaeck}}]{Hibbert1991}
{Hibbert}, A., {Biemont}, E., {Godefroid}, M., \& {Vaeck}, N. 1991, Journal of
  Physics B Atomic Molecular Physics, 24, 3943

\bibitem[{{Hibbert} {et~al.}(1993){Hibbert}, {Biemont}, {Godefroid}, \&
  {Vaeck}}]{Hibbert1993}
{Hibbert}, A., {Biemont}, E., {Godefroid}, M., \& {Vaeck}, N. 1993, \aaps, 99,
  179

\bibitem[{{Hill} {et~al.}(2000){Hill}, {Barbuy}, {Spite}, {Spite}, {Cayrel},
  {Plez}, {Beers}, {Nordstr{\"o}m}, \& {Nissen}}]{Hill2000}
{Hill}, V., {Barbuy}, B., {Spite}, M., {et~al.} 2000, \aap, 353, 557

\bibitem[{{Hubeny} {et~al.}(2021){Hubeny}, {Allende Prieto}, {Osorio}, \&
  {Lanz}}]{Hubeny2021}
{Hubeny}, I., {Allende Prieto}, C., {Osorio}, Y., \& {Lanz}, T. 2021, arXiv
  e-prints, arXiv:2104.02829

\bibitem[{{Keller} {et~al.}(2014){Keller}, {Bessell}, {Frebel}, {Casey},
  {Asplund}, {Jacobson}, {Lind}, {Norris}, {Yong}, {Heger}, {Magic}, {da
  Costa}, {Schmidt}, \& {Tisserand}}]{Keller2014}
{Keller}, S.~C., {Bessell}, M.~S., {Frebel}, A., {et~al.} 2014, \nat, 506, 463

\bibitem[{{Kurucz}(2005)}]{Kurucz2005}
{Kurucz}. 2005, Memorie della Societa Astronomica Italiana Supplementi, 8, 189

\bibitem[{{Kurucz}(2011)}]{Kurucz2011}
{Kurucz}, R.~L. 2011, Canadian Journal of Physics, 89, 417

\bibitem[{{Kurucz}(2018)}]{Kurucz2018}
{Kurucz}, R.~L. 2018, in Astronomical Society of the Pacific Conference Series,
  Vol. 515, Workshop on Astrophysical Opacities, 47

\bibitem[{{Limberg} {et~al.}(2025){Limberg}, {Placco}, {Ji}, {Yao}, {Chiti},
  {Mardini}, {Frebel}, \& {Rossi}}]{Limberg2025}
{Limberg}, G., {Placco}, V.~M., {Ji}, A.~P., {et~al.} 2025, \apjl, 989, L18

\bibitem[{{Luo} \& {Pradhan}(1989)}]{Luo1989}
{Luo}, D. \& {Pradhan}, A.~K. 1989, Journal of Physics B Atomic Molecular
  Physics, 22, 3377

\bibitem[{{M.~Kovalev} {et~al.}(2018){M.~Kovalev}, {S.~Brinkmann},
  {M.~Bergemann}, \& {MPIA IT-department}}]{Kovalev2018}
{M.~Kovalev}, {S.~Brinkmann}, {M.~Bergemann}, \& {MPIA IT-department}. 2018,
  {NLTE MPIA web server, [Online]. Available: {{http://nlte.mpia.de}} Max
  Planck Institute for Astronomy, Heidelberg.}

\bibitem[{{Nordlander} {et~al.}(2019){Nordlander}, {Bessell}, {Da Costa},
  {Mackey}, {Asplund}, {Casey}, {Chiti}, {Ezzeddine}, {Frebel}, {Lind},
  {Marino}, {Murphy}, {Norris}, {Schmidt}, \& {Yong}}]{Nordlander2019}
{Nordlander}, T., {Bessell}, M.~S., {Da Costa}, G.~S., {et~al.} 2019, \mnras,
  488, L109

\bibitem[{{Norris} {et~al.}(2007){Norris}, {Christlieb}, {Korn}, {Eriksson},
  {Bessell}, {Beers}, {Wisotzki}, \& {Reimers}}]{Norris2007}
{Norris}, J.~E., {Christlieb}, N., {Korn}, A.~J., {et~al.} 2007, \apj, 670, 774

\bibitem[{{Osorio} {et~al.}(2020){Osorio}, {Allende Prieto}, {Hubeny},
  {M{\'e}sz{\'a}ros}, \& {Shetrone}}]{Osorio2020}
{Osorio}, Y., {Allende Prieto}, C., {Hubeny}, I., {M{\'e}sz{\'a}ros}, S., \&
  {Shetrone}, M. 2020, \aap, 637, A80

\bibitem[{{Osorio} {et~al.}(2015){Osorio}, {Barklem}, {Lind}, {Belyaev},
  {Spielfiedel}, {Guitou}, \& {Feautrier}}]{Osorio2015}
{Osorio}, Y., {Barklem}, P.~S., {Lind}, K., {et~al.} 2015, \aap, 579, A53

\bibitem[{{Osorio} {et~al.}(2019){Osorio}, {Lind}, {Barklem}, {Allende Prieto},
  \& {Zatsarinny}}]{Osorio2019}
{Osorio}, Y., {Lind}, K., {Barklem}, P.~S., {Allende Prieto}, C., \&
  {Zatsarinny}, O. 2019, \aap, 623, A103

\bibitem[{Placco {et~al.}(2014)Placco, Frebel, Beers, \&
  Stancliffe}]{Placco2014}
Placco, V.~M., Frebel, A., Beers, T.~C., \& Stancliffe, R.~J. 2014, ApJ, 797,
  21

\bibitem[{{Popa} {et~al.}(2023){Popa}, {Hoppe}, {Bergemann}, {Hansen}, {Plez},
  \& {Beers}}]{Popa2023}
{Popa}, S.~A., {Hoppe}, R., {Bergemann}, M., {et~al.} 2023, \aap, 670, A25

\bibitem[{{Santos-Peral} {et~al.}(2025){Santos-Peral},
  {S{\'a}nchez-Bl{\'a}zquez}, {Vazdekis}, {Palicio}, {Knowles}, {Recio-Blanco},
  \& {Allende Prieto}}]{Santos2025}
{Santos-Peral}, P., {S{\'a}nchez-Bl{\'a}zquez}, P., {Vazdekis}, A., {et~al.}
  2025, \aap, 701, A95

\bibitem[{{Sbordone} {et~al.}(2007){Sbordone}, {Bonifacio}, \&
  {Castelli}}]{Sbordone2007}
{Sbordone}, L., {Bonifacio}, P., \& {Castelli}, F. 2007, in IAU Symposium, Vol.
  239, Convection in Astrophysics, ed. F.~{Kupka}, I.~{Roxburgh}, \& K.~L.
  {Chan}, 71--73

\bibitem[{{Starkenburg} {et~al.}(2018){Starkenburg}, {Aguado}, {Bonifacio},
  {Caffau}, {Jablonka}, {Lardo}, {Martin}, {S{\'a}nchez-Janssen}, {Sestito},
  {Venn}, {Youakim}, {Allende Prieto}, {Arentsen}, {Gentile}, {Gonz{\'a}lez
  Hern{\'a}ndez}, {Kielty}, {Koppelman}, {Longeard}, {Tolstoy}, {Carlberg},
  {C{\^o}t{\'e}}, {Fouesneau}, {Hill}, {McConnachie}, \&
  {Navarro}}]{Starkenburg2018}
{Starkenburg}, E., {Aguado}, D.~S., {Bonifacio}, P., {et~al.} 2018, \mnras,
  481, 3838

\bibitem[{{Tomkin} {et~al.}(1992){Tomkin}, {Lemke}, {Lambert}, \&
  {Sneden}}]{Tomkin1992}
{Tomkin}, J., {Lemke}, M., {Lambert}, D.~L., \& {Sneden}, C. 1992, \aj, 104,
  1568

\bibitem[{{Vogt} {et~al.}(1994){Vogt}, {Allen}, {Bigelow}, {Bresee}, {Brown},
  {Cantrall}, {Conrad}, {Couture}, {Delaney}, {Epps}, {Hilyard}, {Hilyard},
  {Horn}, {Jern}, {Kanto}, {Keane}, {Kibrick}, {Lewis}, {Osborne},
  {Pardeilhan}, {Pfister}, {Ricketts}, {Robinson}, {Stover}, {Tucker}, {Ward},
  \& {Wei}}]{Vogt1994}
{Vogt}, S.~S., {Allen}, S.~L., {Bigelow}, B.~C., {et~al.} 1994, in Society of
  Photo-Optical Instrumentation Engineers (SPIE) Conference Series, Vol. 2198,
  Instrumentation in Astronomy VIII, ed. D.~L. {Crawford} \& E.~R. {Craine},
  362

\end{thebibliography}

\end{document}